\documentstyle[twocolumn,prl,aps,floats]{revtex}
\begin{document}
\draft
\wideabs{
\title{Interaction of quasilocal harmonic modes and boson peak in glasses}
\author{D. A. Parshin$^{1,2}$ and C. Laermans$^1$}
\address{$^1$Department of Physics, Katholieke Universiteit Leuven,
Celestijnenlaan 200D, B-3001 Leuven, Belgium}
\address{$^2$St.Petersburg State Technical University, 195251, 
Polytechnicheskaya 29, St.Petersburg, Russia} 
\date{\today}
\maketitle
\begin{abstract}

The direct proportionality relation between the boson peak maximum in
glasses, $\omega_b$, and the Ioffe-Regel crossover frequency for
phonons, $\omega_d$, is established. For several investigated
materials $\omega_b = (1.5\pm 0.1)\omega_d$. At the frequency
$\omega_d$ the mean free path of the phonons $l$ becomes equal to
their wavelength because of strong resonant scattering on quasilocal
harmonic oscillators. Above this frequency phonons {\it cease to
exist}. We prove that the established correlation between $\omega_b$
and $\omega_d$ holds in the general case and is a direct consequence
of bilinear coupling of quasilocal oscillators with the strain field.

\end{abstract}
}

The properties of harmonic vibrational excitations in disordered
media and glasses become now a very active topic of scientific
research~\cite{PM}. Contrary to the quite well established behavior
of electrons in disordered conductors there is no consensus at all
regarding the harmonic vibrations in ordinary glasses. The most
common and challenging of their signatures is the so-called boson
peak observed in numerous experiments in the low-frequency Raman and
inelastic neutron scattering. The physical origin of the peak is
however still a matter of great debates. The common view is that the
solution of this problem is a corner-stone for our veritable
understanding of glassy vibrational dynamics.

The main discussion in the literature involves now the question
whether the harmonic vibrations responsible for the boson peak are
propagating plane waves (phonon like)~\cite{BKM} or localized because
of disorder~\cite{FCV}. The third possibility, which we share in this
paper, is that they are neither propagating waves nor localized but
have a diffusive nature~\cite{AF}.

For the solution of this crucial question quite powerful and
expensive experimental techniques are now in use. First of all these
are Raman experiments themselves. But since visible light due to
energy and momentum conservation laws does not interact properly with
sound-like excitations, one has to use two other possibilities,
namely inelastic X-ray and neutron scattering.

The main difficulty in the inelastic X-ray scattering experiments is
the very high incident photon energy ($\simeq 20$ KeV), and its
relatively small change, of the order of $10^{-7}$.  As a result, the
"X-ray boson peak" is superimposed on the steep wings of strong
elastic line in the forward direction.  Therefore, a serious problem
arises to correctly resolve it and answer the crucial question
whether it changes with the momentum transfer. 

This difficulty explains why recently two independent
groups using the same experimental setup but different fitting
procedures have arrived to completely opposite conclusions about the
propagating character of the excitations in vitreous $SiO_2$ at the boson
peak range~\cite{BKM,FCV}. The situation with inelastic neutron
Brillouin scattering (for {\it small} momentum transfer)
is not much better. 

To solve this problem from our point of view it is necessary to
separate the Brillouin (phonon) lines of the boson peak (as in the
usual light scattering experiments). Then increasing the momentum
transfer $\bf q$ one can observe how the Brillouin line shifts and
broadens approaching the boson peak from the low frequency side.
This scenario implies the existence in the $q$ range $0.1\div 1\,
{\rm nm}^{-1}$ of {\it two peaks}, one from damped phonons (Brillouin
line) and another the boson peak itself.  However such experiments
will hardly appear in a nearest future.

Therefore at this stage of our knowledges any theoretical insight
into the problem would help a lot to establish the true picture of
the harmonic vibrations in glasses. The main idea is that the boson
peak, being a universal glassy property, should be related with other
universal properties of glasses.

In the papers\cite{IKP,PLBL,GPP} the idea was put forward that the
boson peak corresponds to the Ioffe-Regel crossover frequency for
phonons, $\omega_d$. As it was conjectured in\cite{GPP} above
$\omega_d$ phonons {\it cease to exist} as well defined plane wave
excitations. Later this idea was adopted (supported) in many
papers\cite{FCV,FHT}, though taken alone without any theory, it
of course could not explain why actually this correlation occurs. 

The explanation of such a correlation based on the theory of soft
atomic potentials in glasses\cite{KKI} was proposed in\cite{GPP}.
The main idea is that at this frequency the {\it dipole-dipole
interaction} between quasilocal harmonic oscillators renormalizes
their density of states (DOS) from the bare value
$g(\omega)\propto\omega^4$ (for independent oscillators) to the
$g(\omega)\propto\omega$ (for coupled oscillators) behavior. As a
result a boson peak appears in $g(\omega)/\omega^2$ at a frequency
$\simeq\omega_d$. The reason for the DOS transformation is a {\it
strong level repulsion} in the course of the interaction of
oscillators. Above $\omega_d$ the oscillators become {\it
delocalized}~\cite{rem1}.

This correlation was checked indirectly for $As_xSe_{1-x}$
glasses~\cite{PLBL} and excellent agreement was found between the
position of the bump in the reduced specific heat, $C(T)/T^3$
(prototype of the boson peak), and the energy $E_d=\hbar\omega_d$ for
different $x$ compositions. The direct evidence should of course
include the comparison of the boson peak and the Ioffe-Regel
crossover frequency which is explicitly calculated from the mean free
path of phonons $l$. In its turn $l$ can be easily evaluated from the
fit of the thermal conductivity data which are available for many
glasses. The purpose of our paper is just to make such a comparison.

In this paper on the basis of experimental data analysis for
different glasses we have established for a first time a direct
proportionality relation between the position of the boson peak and
the Ioffe-Regel crossover frequency $\omega_d$. At this frequency the
phonon mean free path with respect to the resonant scattering on
quasilocal harmonic oscillators becomes equal to the wave length. To
proceed further we should briefly remind the main theoretical and
experimental results in this field.

It is well known that for many low-temperature properties of glasses
two-level systems (TLS's) and phonons are responsible\cite{WAP}.
However, at higher frequencies it was well established that there is
an excess of {\it additional} low frequency harmonic modes which
dominate the specific heat above a few Kelvin\cite{BND,BPN}.  They
also strongly reduce the phonon mean free path producing the plateau
in the thermal conductivity of glasses\cite{BGGP}.

In Ref.\cite{IKP} it was argued that these excess modes are
quasilocal soft harmonic oscillators (HO) coexisting with TLS's (and
phonons) with density of states increasing as $g(\omega)\propto 
\omega^4$ at moderate frequencies ($W/\hbar < \omega\le\omega_d$, 
see below). As a result with rising the temperature the linear
temperature behavior of the specific heat (TLS's contribution)
changes to $C(T)\propto T^5$ dependence.
Such crossover results in a minimum in $C(T)/T^3$, at some
$T_{\rm min}\approx 0.5-3\,$K\cite{IKP,BGG}.

The quasilocal vibrations 
open also a new effective channel for the phonon scattering.
Their steeply rising DOS, $g(\omega)$, leads to the same
frequency dependence of the inverse mean free path for phonons
because of resonant scattering on these oscillators\cite{BGGP} 
\begin{equation}
l^{-1}_{\rm res, HO} = \frac{\pi}{6\sqrt{2}} 
\frac{C\omega}{v}\left(\frac{\hbar\omega}{W}\right)^3
\propto\omega^4 .  
\label{eq:ert1}
\end{equation}
This dependence looks like elastic Rayleigh scattering from glass
inhomogeneities though the physical mechanism is drastically
different. We believe that in ordinary glasses the Rayleigh
scattering is small in comparison with this contribution.

Except for the sound velocity $v$ there are only two parameters in
(\ref{eq:ert1}). The first one is a characteristic energy $W$
which is related with the position of the minimum in $C(T)/T^3$, 
$W\approx 2kT_{\rm min}$\cite{IKP,BGG}. The
second is a coupling constant $C$ which describes the
relative change of $v$ with temperature 
due to resonance scattering of phonons on TLS's\cite{WAP} 
\begin{equation}
(\Delta v/v)_{\rm res, TLS}=C\ln (T/T_0) .
\label{eq:fdr2}
\end{equation}  

\begin{table}
\caption{Boson peak frequency, $\omega_b$, coupling
constant of longitudinal phonons with TLS's, $C_l$, and $T_{\rm
min}$ for several glasses.} 
\label{tab:hhh}
\begin{center}
\begin{tabular}{llll}
Glass         &$T_{\rm min}$&$\omega_b$&$C_l$\\
              &    (K)      &(${\rm cm}^{-1}$)&($10^{-4}$)\\
\hline
${\rm SiO}_2$     & $^a$2.1  & $^b$52  & $^c$3.1   \\
As$_2$S$_3$      &  $^d$0.8    &   $^f$26  &  $^{c,g}$1.6   \\
B$_2$O$_3$       &  $^a$1   &    $^h$28  &  $^i$2.4   \\
Se                &  $^a$0.6  &  $^j$18  &  $^c$1.2   \\
PS                &  $^k$0.9  &   $^l$17  &  $^c$3.6   \\
GeO$_2$           &  $^m$1.75; 2 &  $^b$45  &  $^n$2.5    \\
LaSF-7            &  2.5 (?)     &  $^o$80  &  $^c$1.2   \\
LiCl$\cdot$7H$_2$O&  $^a$3.3     &  $^p$60  &  $^c$7.2  \\
\end{tabular}
\end{center}
$^a$Ref.\cite{GRB}, $^b$Ref.\cite{RHS}, $^c$Ref.\cite{BM},
$^d$Ref.\cite{RS}, $^f$Ref.\cite{RJN},
$^g$Ref.\cite{LPAH}, $^h$Ref.\cite{MNPS}, $^i$Ref.\cite{PDP},
$^j$Ref.\cite{NP}, $^k$Ref.\cite{ZP}, $^l$Ref.\cite{APSP},
$^m$Ref.\cite{JLW}, $^n$Ref.\cite{LKW}, $^o$Ref.\cite{PTP},
$^p$Ref.\cite{BG}.
\end{table}

Both experimental values of $C$ and $T_{\rm min}$ are well known for
many glasses. It gave the unique possibility, without any fitting
parameters using (\ref{eq:ert1}), to successfully reproduce the
correct value of the thermal conductivity plateau for vitreous
$SiO_2$ and $Se$\cite{BGGP} and for some other glasses as
well\cite{RB}. From this point of view one can consider
Eq.(\ref{eq:ert1}) to be in a good agreement with the thermal
conductivity data.

Using (\ref{eq:ert1}) one can easily obtain the expression for the
Ioffe-Regel crossover frequency $\omega_d$\cite{PLBL,GPP} 
\begin{equation}
\hbar\omega_d=0.75 WC^{-1/3} .
\label{eq:hg7l}
\end{equation}  
For example for $v-SiO_2$, $C_l=3.1\times 10^{-4}$, $W/k=4.2\,$K and
$\hbar\omega_d/k \approx 47\,$K. 

Now we can proceed to the main aim of our paper and check whether
there exists a correlation between the boson peak frequency
$\omega_b$ and $\omega_d$ determined by (\ref{eq:hg7l}). For that
purpose we collected in the Table~\ref{tab:hhh} all the necessary
experimental parameters $T_{\rm min}$, $C_l$ and $\omega_b$ for
several glasses.  The two different values of $T_{\rm min}$ for
$a-GeO_2$ correspond to two different samples investigated
in~\cite{JLW}. The value of $T_{\rm min}$ for LaSF-7 glass is a
reasonable guess value (we did not find it in a literature).

Now using the relation $W=2kT_{\rm min}$ and
Eq.~\ref{eq:hg7l} we can calculate $\omega_d$ for longitudinal
phonons and compare it with 
$\omega_b$. The result of this comparison is shown on
Fig.1. One can see from the figure that all the data
lies near a straight line which has the slope equal to $1.53$.
It means that within the experimental accuracy there is
a direct proportionality relation between the boson peak maximum
$\omega_b$ and the Ioffe-Regel crossover frequency for phonons
$\omega_d$.  

In what follows we are going to prove that this correlation
results in fact from the {\it bilinear coupling} between quasilocal
oscillators and the deformation field $\varepsilon$  
\begin{equation}
{\cal H}_{\rm int}=\Lambda x\varepsilon 
\label{eq:dfg4}
\end{equation}    
proposed in\cite{BGGP}. In this sense it can be regarded as a rather
general rule. 

With the use of (\ref{eq:dfg4}) one can easily show that
the elastic interaction between two oscillators has a dipole-dipole
character 
\begin{equation}
V_{\rm int}\simeq \frac{\Lambda^2}{\rho v^2 r_{ij}^3}x_ix_j 
\label{eq:wxao}
\end{equation}  
with $\rho$ being the mass density of the glass.

Suppose that the DOS of noninteracting
oscillators is $g(\omega)$. Then making use of (\ref{eq:dfg4}) and
the Golden rule one can calculate in the usual way the phonon mean
free path due to the resonance scattering on these oscillators 
\begin{equation}
l^{-1}_{\rm res,HO}=\frac{\pi\Lambda^2}{2M\rho v^3}g(\omega)
\label{eq:df7f}
\end{equation}   
where $M$ is an oscillator mass. It is important that this 
scattering has the same frequency dependence as the bare density 
$g(\omega)$. The Ioffe-Regel crossover frequency $\omega_d$ can be
estimated from the equation 
\begin{equation}
\left.\frac{g(\omega)}{\omega}\right|_{\omega_d}
\simeq \frac{M\rho v^2}{\Lambda^2} .
\label{eq:dfr4}
\end{equation}  

In order to determine a position of the boson peak we should consider
the interaction between resonant oscillators. The concentration 
of such oscillators with frequencies $\omega$ lying in a small
interval $\delta\omega$ around $\omega$ is given by the product 
$g(\omega)\delta\omega$. Therefore the non-diagonal transition matrix
element of the coupling between the two oscillators (\ref{eq:wxao})
is equal 
\begin{equation}
\Delta J\equiv
\left< n_{i+1},n_j|V_{\rm int}|n_i,n_{j+1}\right> \simeq 
\frac{\hbar\Lambda^2 g(\omega)}{2M\rho v^2\omega}\delta\omega .
\label{eq:dp2d}
\end{equation}    
The physical meaning of this quantity is that it gives a characteristic
value for a level repulsion of two resonant oscillators because of their
interaction.

If $\Delta J < \hbar\delta\omega$ the level repulsion is weak and the
interaction does not change $g(\omega)$.
If $\Delta J > \hbar\delta\omega$ the level repulsion is strong and
renormalizes the original density\cite{GPP}. The new DOS
can be found dividing the oscillator concentration in the
interval by the characteristic value of their level repulsion 
\begin{equation}
\widetilde g(\omega) \simeq \frac{g(\omega)\delta\omega}
{\Delta J/\hbar}\simeq \frac{M\rho v^2}{\Lambda^2}\omega .
\label{eq:df4q}
\end{equation}  
We see that the result is a {\it linear} function of $\omega$ which
depends only on material parameters of the glass but does not depend
on the bare density $g(\omega)$. 

In the case when the reduced bare DOS $g(\omega)/\omega^2$ increases
with $\omega$ we will have a boson peak at some crossover frequency
from weak to strong coupling. This frequency is determined by the
equation $\Delta J = \hbar\delta\omega$. One can easily see that this
equation coincides with equation (\ref{eq:dfr4}) for $\omega_d$.
This explains why $\omega_d$ correlates with the position of the
boson peak in glasses.

The resulting linear $\omega$ dependence of vibrational DOS above the
boson peak is a very important and general feature of a system of
interacting harmonic oscillators with bilinear dipole-dipole coupling
(\ref{eq:wxao}). It is a direct consequence of linear $\omega$
dependence of $|x_{n,n\pm 1}|^{-2}=2M\omega/\hbar$ and therefore can
be regarded as universal.  It holds for any bare $g(\omega)$ of
quasilocal harmonic modes which rises with frequency faster then
$\omega^2$. It was observed in experiments on inelastic neutron
scattering for amorphous polymers\cite{UB}, $a-SiO_2$~\cite{WBD} and
$a-GeSe_2$\cite{CCD}.

The picture described above implies delocalization of the harmonic
excitations above the boson peak though they are not supposed to be
plane waves. This is also consistent with existing numerical
results. The fact that the Ioffe-Regel criterion does not mean
localization for vibrational excitations in disordered system was
acknowledged in the computer analysis of the percolation
network\cite{SZZ}. It was found that vibrational modes above the
Ioffe-Regel crossover give substantial contribution to the heat
transfer and have some diffusive nature.

Similar results were obtained in\cite{JVKR} for $a-SiO_2$ where all
modes with $\hbar\omega$ between $5-110\, {\rm meV}$ were found to be
delocalized (just to the right of the boson peak). For amorphous
silicon it was found\cite{FKAW} that only 3\% of vibrational modes
(above $70$ meV) are localized and do not participate in the heat
transfer. The remaining 97\% of the modes are delocalized and the
majority of them (93\%) are not plane waves.

In numerical experiments\cite{SL} the truly localized modes were
discovered at the very end of the frequency spectrum. It is
worthwhile to mention that at low frequencies the calculated DOS has
a part with linear $\omega$ dependence just in the region of
delocalized states. The same results were obtained for amorphous
selenium\cite{OS}. The modes with frequencies above $0.5\, {\rm THz}$
were delocalized (again just above the boson peak for this glass).

To conclude, we have established for the first time the correlation
between the boson peak frequency and the Ioffe-Regel crossover
frequency. We proved that this correlation occurs because of the 
bilinear coupling between sound waves and quasilocal harmonic
vibrations in glasses. 

One of the authors (D.A.P.) gratefully acknowledge pleasant
hospitality and financial support of the Catholic University of
Leuven where this work was done. One of the authors
(C.L.) thanks the Belgian FWO for financial support. 


\centerline{Figure caption}

Fig.1 Position of the boson peak $\omega_b$ versus the Ioffe-Regel
crossover frequency $\omega_d$ for several glasses.


\begin{references}
\bibitem{rem1} This explains subscript letter "d" in $\omega_d$.
By term of "delocalization" we understand that being excited harmonic
oscillator most probably transfer its energy not to a phonon (in such
a case it would be a quasilocal oscillator) but to another oscillator.
\end{references}
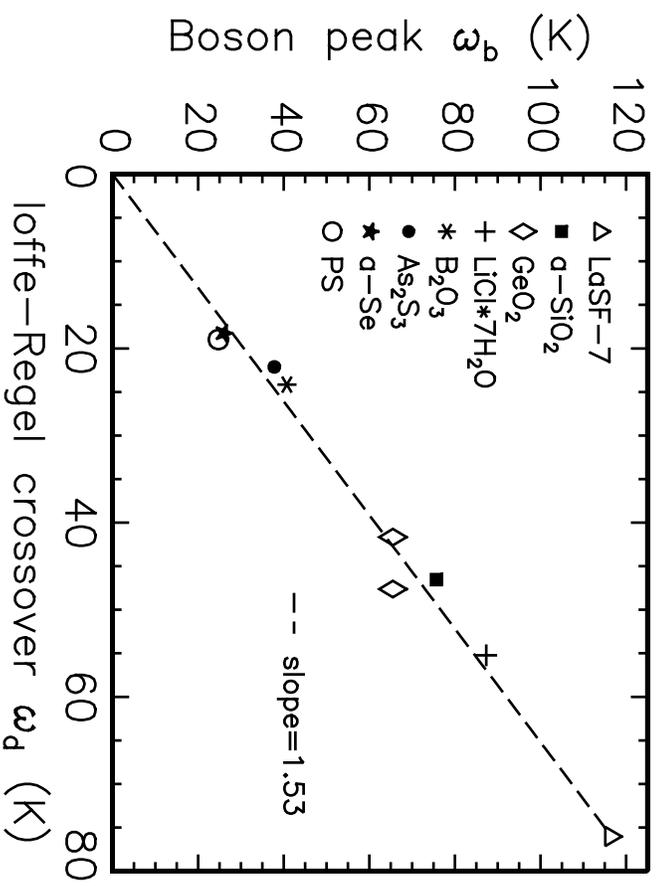
\end{document}